\documentclass[12pt]{book}
\usepackage[dvips]{graphicx,color}
\usepackage{makeidx,tsukuba}

\makeauthorindex
\makeindex

\begin{document}
\BookTitle{\itshape The 28th International Cosmic Ray Conference}
\CopyRight{\copyright 2003 by Universal Academy Press, Inc.}
%\tableofcontents
\pagenumbering{arabic}

\chapter{Results from the BAIKAL Neutrino Telescope}
\noindent
\author{
V.Ayutdinov$^1$,
V.Balkanov$^1$,
I.Belolaptikov$^4$,
L.Bezrukov$^1$,
N.Budnev$^2$,
A.Chensky$^2$,
D.Chernov$^3$,
I.Danilchenko$^1$,
Zh.-A.Dzhilkibaev$^1$,
G.Domogatsky$^1$,
A.N.Dyachok$^2$,
O.Gaponenko$^1$,
O.Gress$^2$,
T.Gress$^2$,
A.Klabukov$^1$,
A.Klimov$^8$,
S.Klimushin$^1$,
K.Konischev$^4$,
A.Koshechkin$^1$,
V.Kulepov$^6$,
L.Kuzmichev$^3$,
Vy.Kuznetzov$^1$,
B.Lubsandorzhiev$^1$,
S.Mikheyev$^1$,
M.Milenin$^6$,
R.Mirgazov$^2$,
N.Moseiko$^3$,
E.Osipova$^3$,
A.Panfilov$^1$,
G.Pan'kov$^2$,
L.Pan'kov$^2$,
Yu.Parfenov$^2$,
A.Pavlov$^2$,
E.Pliskovsky$^4$,
P.Pokhil$^1$,
V.Polecshuk$^1$,
E.Popova$^3$,
V.Prosin$^3$,
M.Rosanov$^7$,
V.Rubtzov$^2$,
Yu.Semeney$^2$,
B.Shaibonov$^1$,
Ch.Spiering$^5$,
O.Streicher$^5$,
B.Tarashanky$^2$,
R.Vasiliev$^4$,
E.Vyatchin$^1$,
R.Wischnewski$^5$,
I.Yashin$^3$,
V.Zhukov$^1$ \\
{\it (1) Institute for Nuclear Research, Moscow, Russia, 
(2) Irkutsk State University, Irkutsk,Russia, 
(3) Skobeltsyn Institute of Nuclear Physics  MSU, Moscow, Russia,
(4) Joint Institute for Nuclear Research, Dubna, Russia,
(5) DESY--Zeuthen, Zeuthen, Germany,
(6) Nizhni Novgorod State Technical University, Nizhni Novgorod, Russia,
(7) St.Peterburg State Marine University, St.Petersburg, Russia,
(8) Kurchatov Institute, Moscow, Russia} \\
\vspace{-2mm}
}

\section*{Abstract}

We review the present status of the Baikal Neutrino Project,
present updated results on the search for
high energy extraterrestrial neutrinos,
fast magnetic monopoles and neutrinos induced by WIMP annihilation
in the center of the Earth and compare the recorded atmospheric 
neutrino flux to predictions.
%We describe the moderate upgrade of NT-200
%toward $\sim$10 Mton scale detector NT-200$+$.

\vspace{0.2cm}
\section{Introduction}

\vspace{-0.1cm}
The Baikal Neutrino Telescope NT-200  is operated in Lake 
Baikal, Siberia,  at a depth of \mbox{1.1 km}. 
A description of the detector as well
as physics results from data collected in 1996 and 1998
(70 and 234 live days, respectively) 
have been presented elsewhere [1-3]. Here we present
new limits including data taken in 1999 (268 live days).
Data taken in 2000 are presently being analyzed.
We also describe  NT-200$+$ -- an upgrade of NT-200 by 
three sparsely instrumented distant outer strings
which increase the fiducial volume for high energy cascades
to a few dozen Mtons. A prototype string of 140 m length
with 12 optical modules was deployed in March 2003, and
electronics, data acquisition
and calibration systems for NT-200$+$ have been tested.

\vspace{-0.3cm}
\section{Atmospheric Muon Neutrinos as Calibration Tool}
\vspace{-0.3cm}
The clearest signature of neutrino induced events is a muon 
crossing the detector from below. 
Track reconstruction algorithms as well as 
background rejection have been described elsewhere [1].
Since on the one side
the energy threshold for this particular analysis (15-20 GeV)
is higher than in underground detectors, and 
on the other side NT-200 is
much smaller than Amanda, event rates of
atmospheric neutrinos are small compared to these experiments.
Nevertheless, atmospheric neutrinos serve as an important
calibration tool and
demonstrate the understanding of the detector performance.
The 
data set yields 84
upward going muons. The MC simulation of upward muon tracks
due to atmospheric neutrinos
gives 80.5 events.
The angular distribution for both experiment and simulation 
as well as the skyplot of 
upward muons 
are shown in Fig. 1.
%%%%%%%%%%%%%%%%%%%%%%%%%%%%%%%%%%%%%%%%%%%%%%%%%%%%%%%%%%%%%%%%%%%
\begin{figure}[htbp]
  \vspace{-0.2cm}
\includegraphics[width=.43\textwidth]{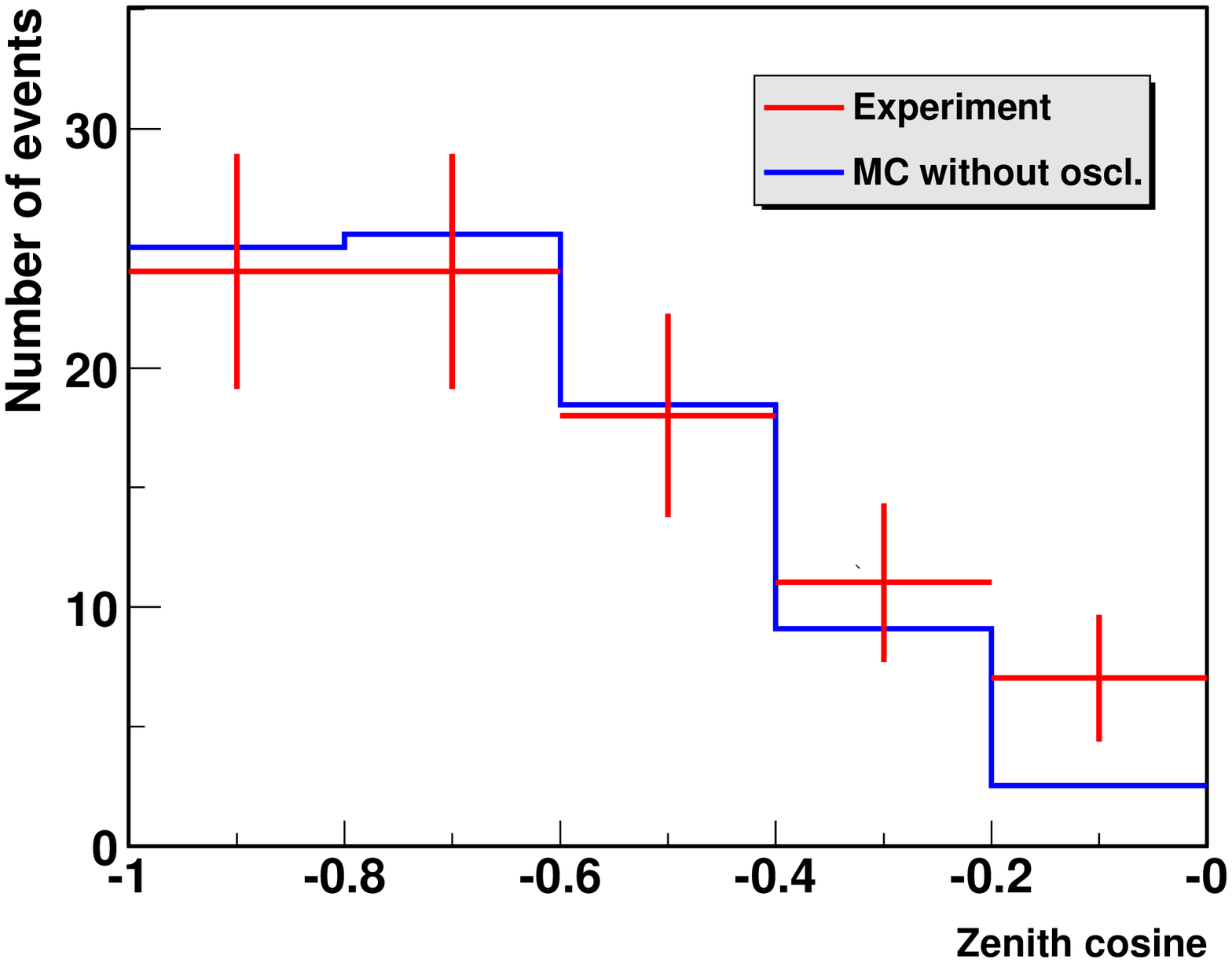}
\hfill
\vspace{-4mm}
\includegraphics[width=.5\textwidth, height=4.5cm]
{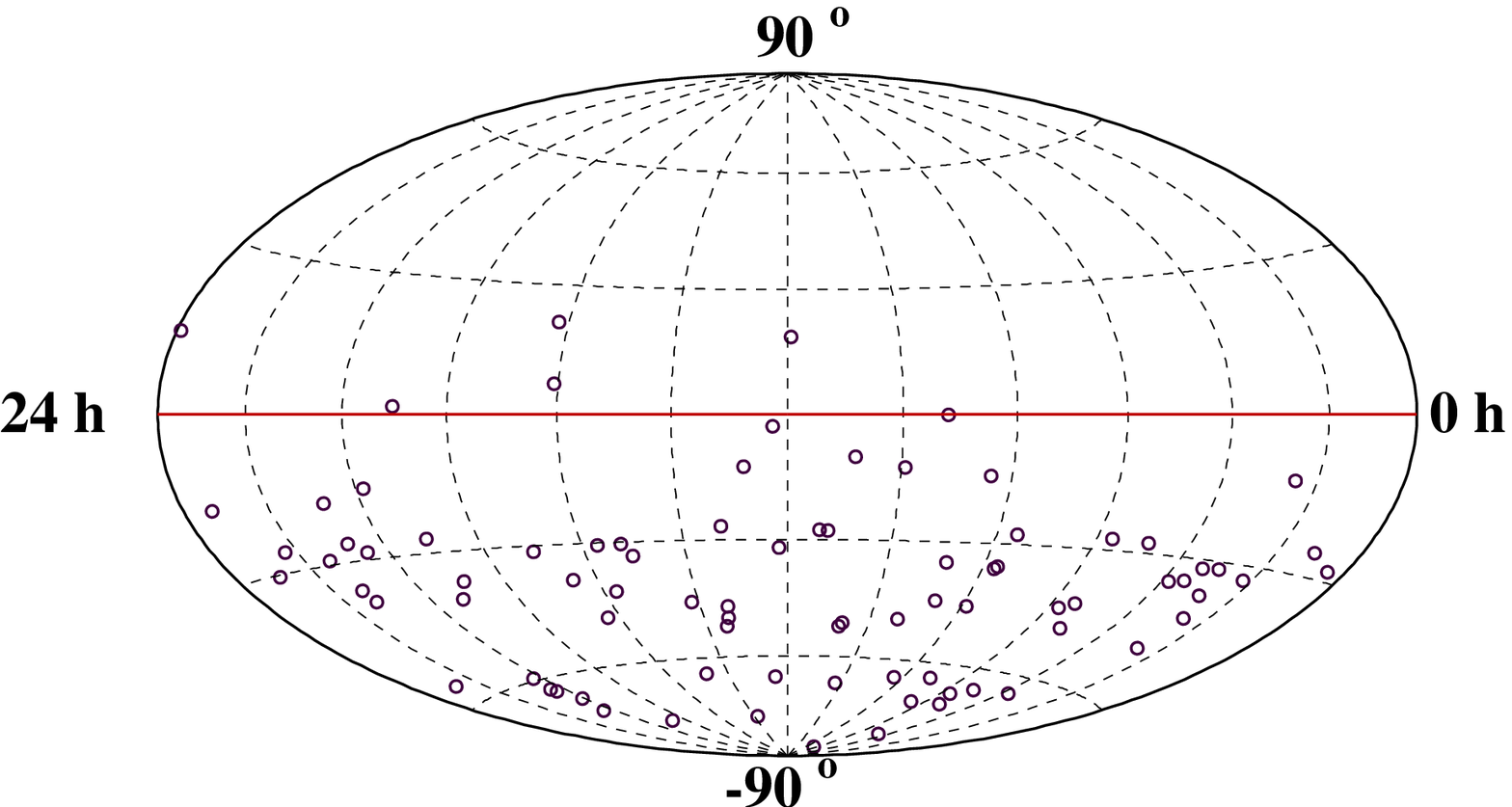}
\\
  \vspace{-0.4cm}
\caption{Left: Angular distribution of experimental events 
and MC data. 
Right: Skyplot (equatorial coordinates) of neutrino events. 
}
\label{mons2}
\end{figure}
%%%%%%%%%%%%%%%%%%%%%%%%%%%%%%%%%%%%%

\vspace{-0.6cm}
\section{Search for Neutrinos from WIMP Annihilation}
\vspace{-0.3cm}
The search for WIMPs with the Baikal
neutrino telescope is based on a possible signal of
nearly vertically upward going muons, exceeding
the flux of atmospheric neutrinos [2]. With no
significant excess observed, we derive improved 
upper limits on the flux of muons from the direction
of the center of Earth related to WIMP annihilation.
Note that the threshold of 8-10 GeV for this analysis
is lower than that for atmospheric neutrinos spread
across the full lower hemisphere (see above).
Fig.2 compares our new limits to those obtained by other 
experiments (see [4] and references given in [2]).

\begin{figure}[htbp]
\vspace{-0.9cm}
\includegraphics[width=.45\textwidth]{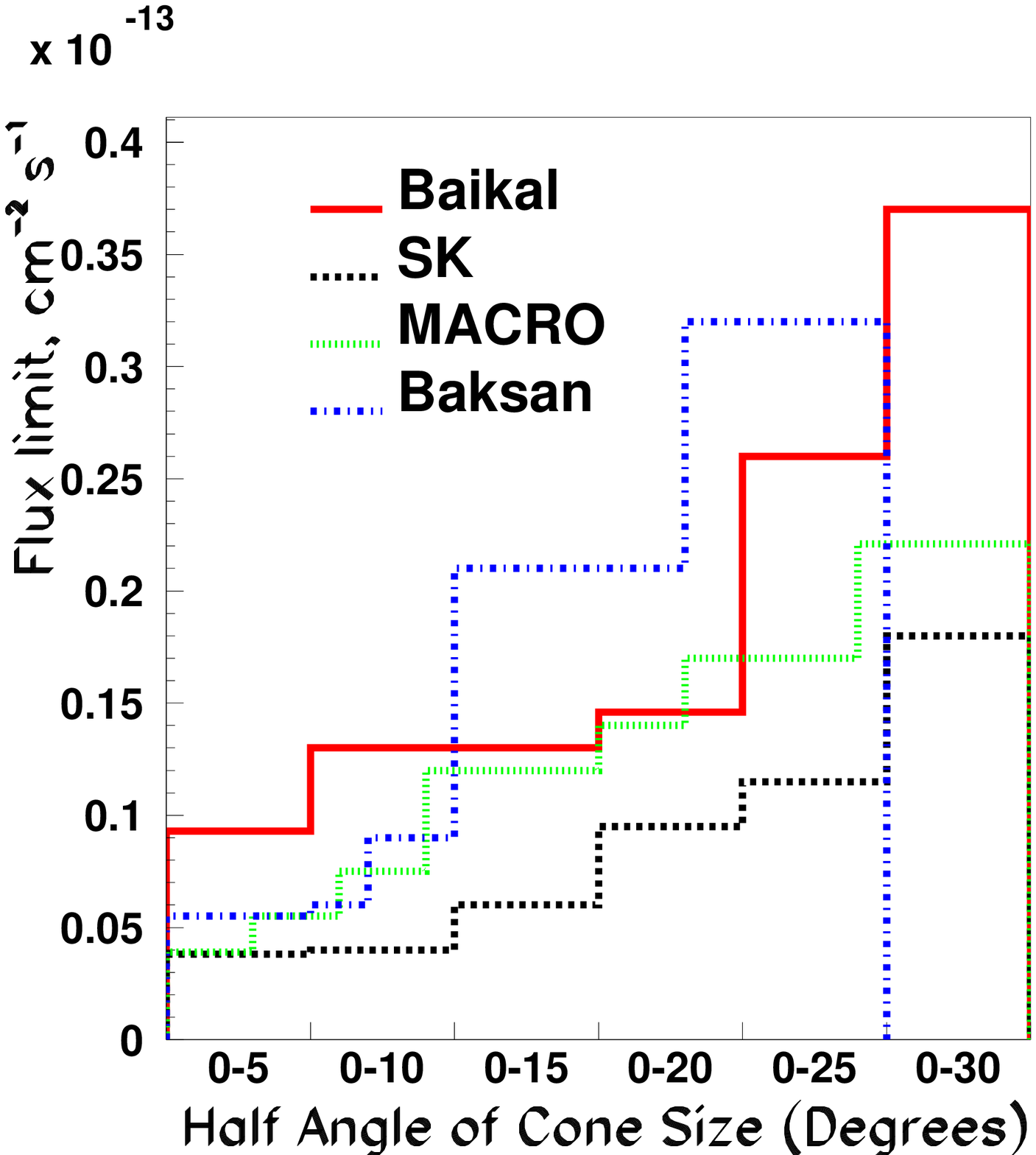}
\hfill
\includegraphics[width=.45\textwidth]{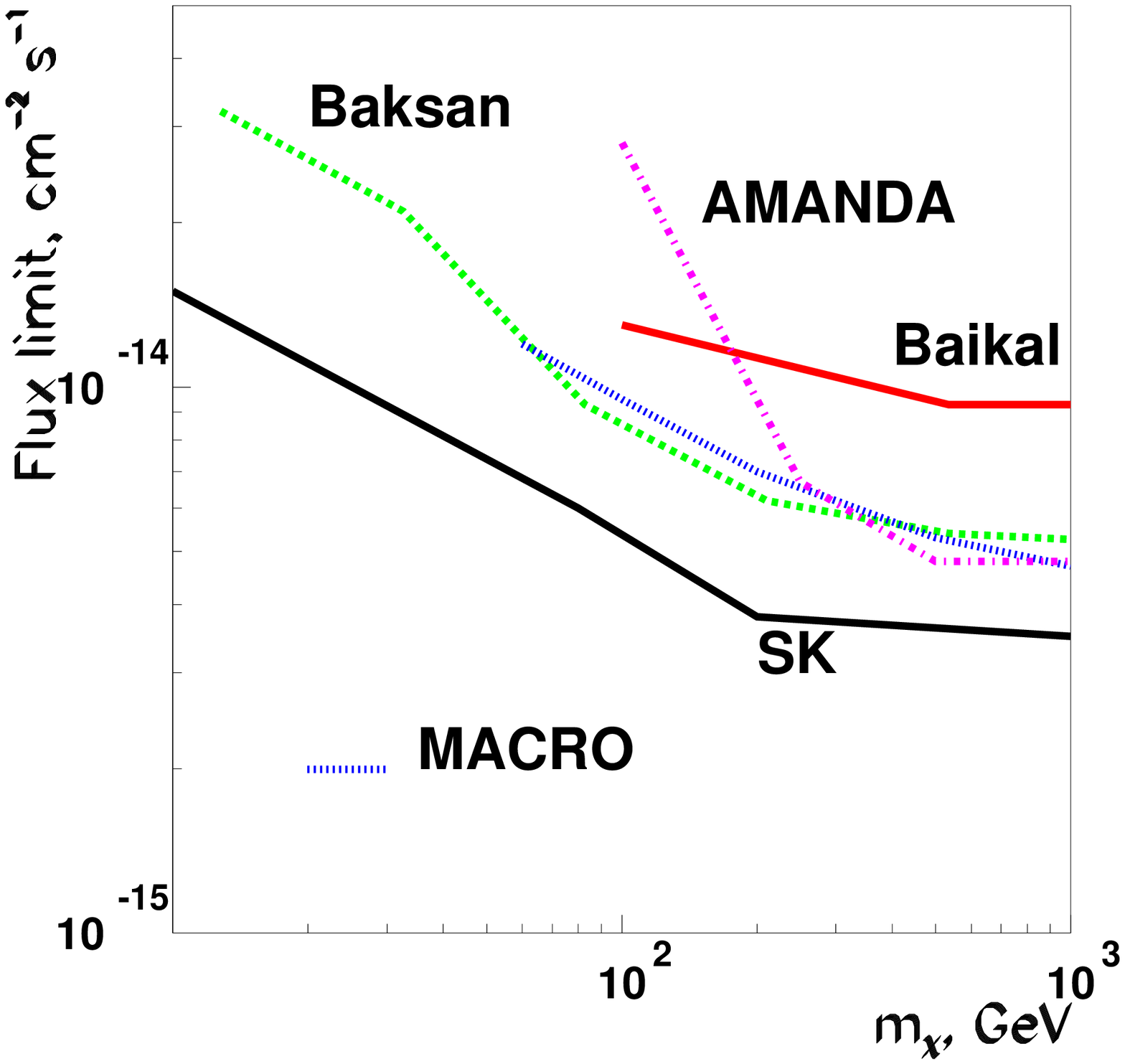}
\\
  \vspace{-0.5cm}
  \caption{Left: Limits on the excess muon flux from the center of 
  the Earth versus half-cone of the search angle. Right: 
  Flux limits as a function of WIMP mass.
}
\label{mons2}
\end{figure}

\vspace{-0.1cm}
\section{Search for Relativistic Magnetic Monopoles}

\vspace{-0.3cm}
Events due to relativistic monopoles ($\beta >$ 0.75)
are distinguished by their high light output, allowing
identification of events beyond the geometrical boundaries
of the detector.
The search strategy has been described in [2]. An improved
analysis including data from 1996 to 1999 yields a limit
about a factor of two below the limit published earlier.
This limit is compared to those from
other experiments (see [5] and references in [2]) in Fig. 3.
%%%%%%%%%%%%%%%%%%%%%%%%%%%%%%%%%%%%%%%%%
\begin{figure}[htbp]
\vspace{-0.1cm}
\includegraphics[width=0.48\textwidth,height=5.5cm]{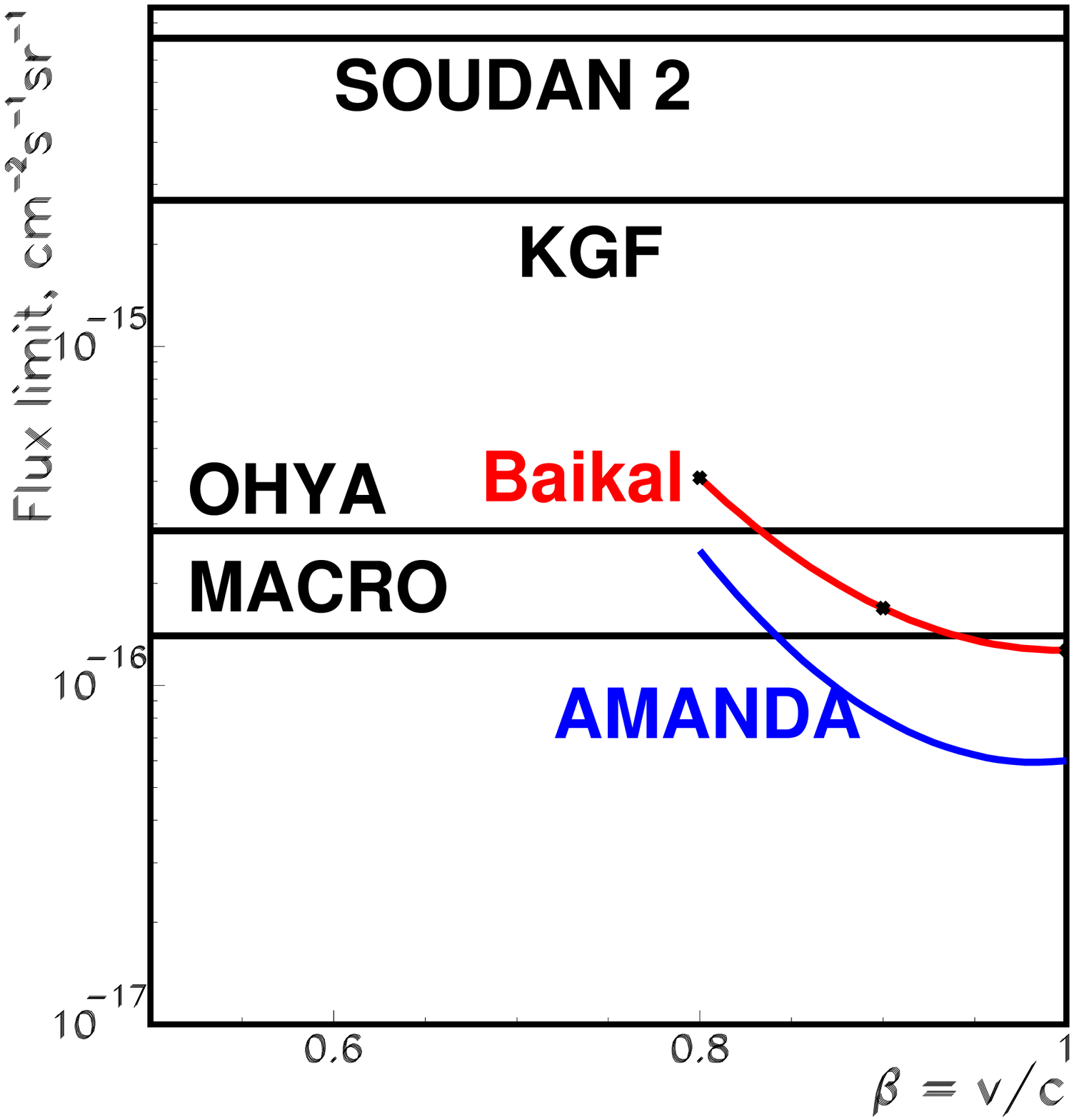}
\hfill
\parbox[b]{.5\textwidth}                
{{Fig. 3.}: Upper limits on the flux of fast monopoles
obtained in different experiments.}
\\
\end{figure}
%%%%%%%%%%%%%%%%%%%%%%%%%%%%%%%%%%%%%%%%%%%%%%%%%%%%

\vspace{-0.7cm}
\section{A Search for Extraterrestrial High Energy Neutrinos}
\vspace{-0.3cm}
The BAIKAL survey for high energy neutrinos searches
for bright cascades produced at the neutrino interaction
vertex in a large volume around the neutrino telescope [3].
Lack of significant light scattering allows to monitor a 
volume exceeding the geometrical volume by an order of magnitude.
This results in sensitivities of NT-200 comparable
to those of the much larger Amanda-B10 detector. 
The background to this search are bright bremsstrahlung flashes
along downward muons passing far outside the array.
The method has been described in [3].

Candidate events do not show a statistically
significant excess of hit multiplicity compared to the
simulated background from atmospheric muons.
Assuming an $E^{-2}$ shape of the neutrino spectrum and a
flavor ratio  $\nu_e:\nu_{\mu}:\nu_{\tau}=1:1:1$,  
the new, preliminary 90\% C.L. upper limit is
$\Phi_{\nu_e}E^2=4\cdot10^{-7} 
\mbox{cm}^{-2}\mbox{s}^{-1}\mbox{sr}^{-1}\mbox{GeV}$,
about twice below previous results [3]. 
The preliminary limit on $\tilde{\nu_e}$ at the W - resonance 
energy is: $\Phi_{\tilde{\nu_e}} \leq 5.4 \times 
10^{-20} 
\mbox{cm}^{-2}\mbox{s}^{-1}\mbox{sr}^{-1}\mbox{GeV}^{-1}.$
These limits do not yet
include the effect of systematic uncertainties.
Fig.4 (left) shows the experimental upper limits [3] as well as 
theoretical limits 
obtained by Berezinsky (B), by Waxman and Bahcall (WB), 
by Mannheim et al.(MPR), 
and predictions for neutrino fluxes from Stecker and 
Salamon (SS) and Protheroe (P).

Fig.4 (right) shows NT-200+ with its three additional outer 
(plus one possible central) strings.
It will allow a much better vertex identification and hence
a significantly more precise
{\it measurement} of cascade energy in a volume around NT-200.
The sensitivity of NT-200+ to high energy
cascades will be four times better than that of NT-200, with
a moderate 20 percent increase of optical modules only.

%%%%%%%%%%%%%%%%%%%%%%%%%%%%%%%%%%%%%%
\begin{figure}[htb]
  \begin{center}
\vspace{-0.5cm}
\includegraphics[width=7.0cm]{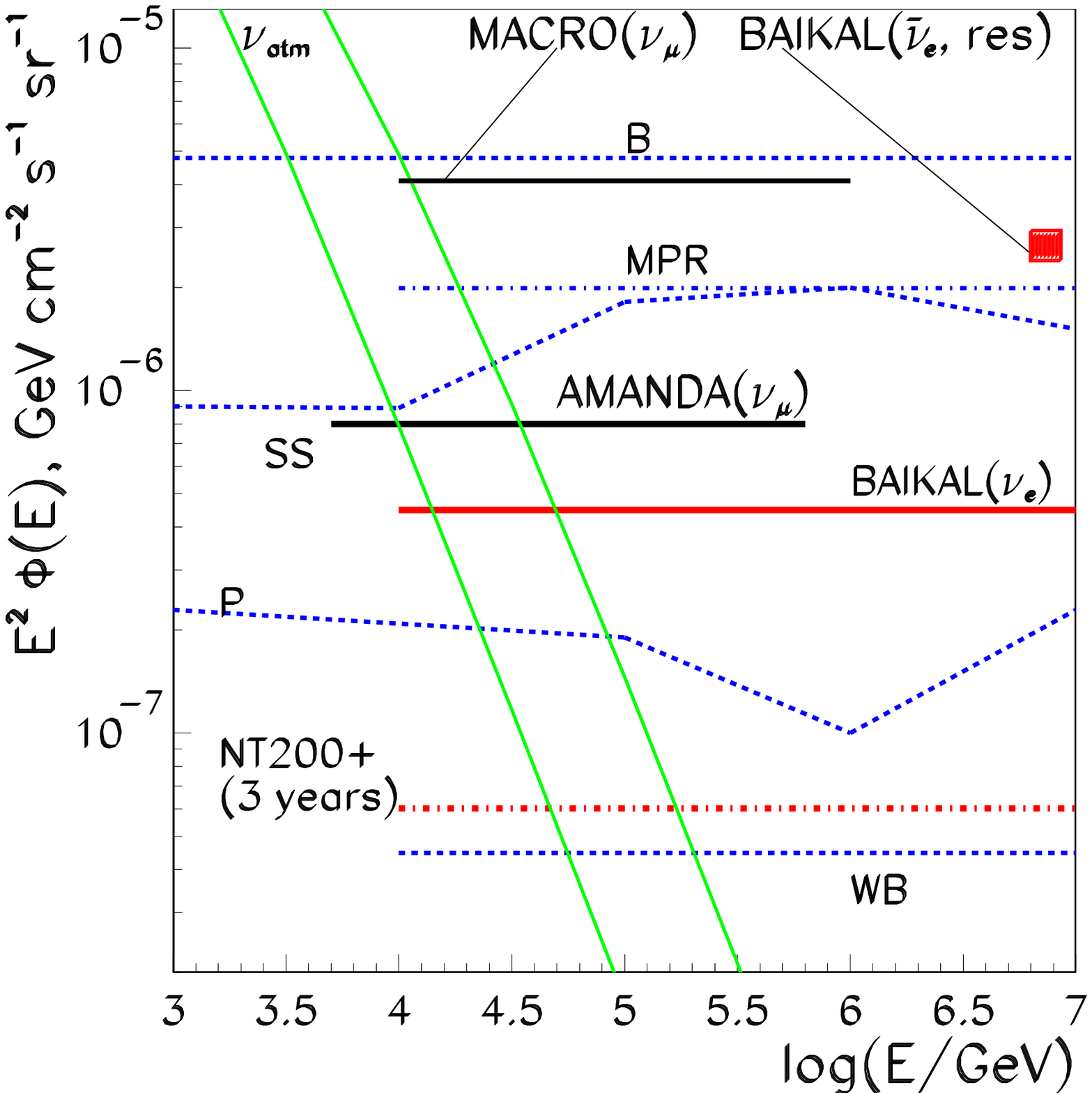}
\hfill
\includegraphics[width=6.2cm]{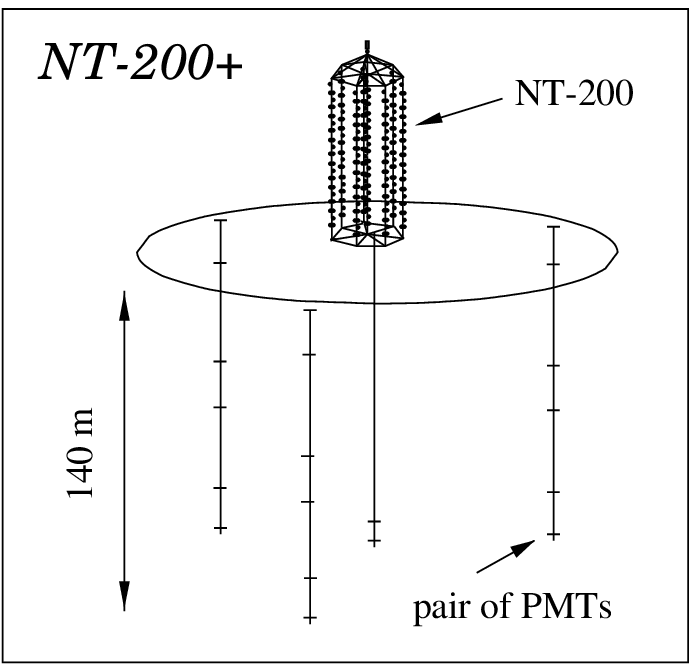}
\end{center}
\caption{Left: Experimental upper limits on the neutrino 
fluxes as well as flux 
predictions in different models of neutrino sources (see text).
Right: The NT-200+ configuration.
}
\end{figure}
%%%%%%%%%%%%%%%%%%%%%%%%%%%%%%%%%%%%%%%%%%%%%%%%%%%%%%%%%%%%

%\bigskip
\vspace{-0.5cm}
{\it This work was supported by the Russian Ministry of Research,the German 
Ministry of Education and Research and the Russian Fund of Basic 
Research ( grants } \mbox{\sf 03-02-31004}, \mbox{\sf 02-02-17031}, 
\mbox{\sf 02-07-90293} {\it and} \mbox{\sf 01-02-17227}),
{\it Grant of President of Russia} \mbox{\sf NSh-1828.2003.2} 
{\it and by the Russian Federal Program ``Integration'' (project no.} 
\mbox{\sf E0248}).

\vspace{-0.3cm}
\section{References}

\vspace{-0.3cm}
%\vspace{\baselineskip}
\re
1.\ Belolaptikov I.\ et al.\ 1997, Astr.Ph. 7, 263;
and Astr.Ph. 12 (1999) 75.
\re
2.\ Balkanov V.\ et al.\ 2001, Nucl.Ph.(Proc.Sup.) B91, 438.
\re
3.\ Balkanov V.\ et al., astro-ph/0211571.
\re
4.\ Habig A.\ et al.\ 2001, Proc.of the XVII ICRC, 1558.
\re
5.\ Ambrosio M.\ et al.\ 2002, Eur.Ph.J. C25, 511.

\endofpaper

\end{document}